# Are Two of the Neptune Trojans Dynamically Unstable?

Jonathan Horner [1], Patryk Sofia Lykawka [2]

[1] *Department of Astrophysics, School of Physics, University of New South Wales, Sydney, NSW 2052, Australia, contact e-mail: j.a.horner@unsw.edu.au*
[2] *Astronomy Group, Faculty of Social and Natural Sciences, Kinki University, Shinkamikosaka 228-3, Higashiosaka-shi, Osaka, 577-0813, Japan*

**Summary:** The Neptune Trojans are the most recently discovered population of small bodies in the Solar System. To date, only eight have been discovered, though it is thought likely that the total population at least rivals that of the asteroid belt. Their origin is still the subject of some debate. Here, we detail the results of dynamical studies of two Neptune Trojans, 2001 QR$_{322}$ and 2008 LC$_{18}$. We find that both objects lie very close to boundaries between dynamically stable and unstable regions, with a significant probability that either or both of the objects are actually unstable on timescales of a few hundred million years. Such instability supports the idea that at least these two Neptune Trojans are dynamically captured objects, rather than objects that formed in situ. This does not, however, rule out the possibility that these two objects were captured during Neptune's proposed post-formation migration, and have remained as Trojans ever since.

**Keywords:** Neptune Trojans, Solar System formation, Solar System evolution, Centaurs, Comets, 2001 QR$_{322}$, 2008 LC$_{18}$, Dynamical Methods

## Introduction

The Neptune Trojans are the most recent addition to the panoply of Solar System small body populations. Though the possible existence of Neptunian Trojans has been debated for decades [1], the first, 2001 QR$_{322}$, was discovered just ten years ago. Since that discovery, a further seven Neptunian Trojans have been discovered (2004 UP$_{10}$, 2005 TN$_{53}$, 2005 TO$_{74}$, 2006 RJ$_{103}$, 2007 VL$_{305}$, 2008 LC$_{18}$ and 2004 KV$_{18}$). Based on the discovery of the first four of these objects, it was estimated that the total population of Neptunian Trojans might well outnumber that of the Jovian Trojans (of which almost 5,000 are known of a population estimated to exceed that of the main asteroid belt[1]) [2,3].

---

[1] The asteroid belt is a broad region between the orbits of Mars and Jupiter that is populated by a large number of primarily rocky bodies. The great majority of the objects contained within the belt move on orbits that are dynamically stable on timescales comparable to, or longer than, the age of the Solar System. However, a continual flux of material diffuses from the belt to the inner Solar System, repeatedly re-populating the near-Earth asteroid population (e.g. [29]). Planetary Trojans are objects trapped in 1:1 mean-motion resonance with a given planet. Typically, Trojans follow "tadpole" shaped paths as they librate around their host planet's L4 and L5 Lagrange points. Less frequently, Trojans can be found following "horseshoe" shaped paths that allow their libration to encompass both the L4 and L5 points of their host planet. Typically, such horseshoe orbits are significantly less stable than tadpole orbits, and as such, the great majority of known Trojans within the Solar System are "tadpole Trojans". For an elegant representation of tadpole and horseshoe orbits, we direct the interested reader to figures 1 and 2 of [37].

As is the case with their better known brethren, the Jovian Trojans, the Neptunian Trojan[2] population displays an orbital distribution that is strikingly different from that which might be expected under the assumption that they formed in situ from a dynamically cold disk of gas and dust (i.e. a disk in which the particles move on orbits with typically very small orbital inclinations and eccentricities). Rather than displaying typically dynamically cold orbits, the eight bodies known range widely in both orbital eccentricity and inclination, with just two having orbital inclinations of less than five degrees (2001 $QR_{322}$ and 2004 $UP_{10}$). Indeed, three of the population move on orbits with inclinations in the range 25 – 30 degrees (2005 $TN_{53}$, 2007 $VL_{305}$ and 2008 $LC_{18}$), whilst five possess eccentricities exceeding 0.05. The most eccentric member, 2004 $KV_{18}$, has a remarkable $e = 0.1842$! Table 1 lists the orbital characteristics and estimated sizes of currently known Neptune Trojans.

| Prov. Des. | $L_n$ | $a$ (AU) | $e$ | $i$ (°) | $D$ (km) |
|---|---|---|---|---|---|
| 2001 $QR_{322}$ | 4 | 30.396 | 0.0306 | 1.32 | 100-200 |
| 2004 $UP_{10}$ | 4 | 30.318 | 0.0323 | 1.43 | 50-100 |
| 2005 $TN_{53}$ | 4 | 30.285 | 0.0678 | 24.96 | 50-100 |
| 2005 $TO_{74}$ | 4 | 30.296 | 0.0524 | 5.24 | 50-100 |
| 2006 $RJ_{103}$ | 4 | 30.201 | 0.0287 | 8.16 | 100-200 |
| 2007 $VL_{305}$ | 4 | 30.197 | 0.0684 | 28.08 | 80-150 |
| 2008 $LC_{18}$ | 5 | 29.937 | 0.0838 | 27.57 | 80-150 |
| 2004 $KV_{18}$ | 5 | 30.126 | 0.1842 | 13.61 | 50-100 |

Table 1: The best-fit orbits of the eight known Neptune Trojans, taken from the AstDys orbital database on 19th October 2011. Here, *Prov. Des.* gives the provisional designation of each of the objects. $L_n$ details the Lagrange point in the Neptune-Sun system about which the Trojan is librating, whilst *a* denotes the semi-major axis, *e* the eccentricity, and *i* the inclination of the orbit. In addition, *D* gives the range of equivalent diameters of the objects in km, assuming that they have albedos of 0.05 (upper estimate) or 0.20 (lower estimate).

On the basis of this somewhat surprising orbital distribution, it is clearly important to obtain a better understanding of the formation, evolution and dynamical behaviour of objects in the Neptunian Trojan clouds. The initial studies that considered the formation of the Neptunian Trojans assumed them to be objects that had formed and were already trapped in resonance with Neptune by the time the planet's formation was complete. Such models share the common property that the objects formed from a dynamically cold disk of material, and were then either captured by the giant planet as it accreted (through a gravitational "pull-down" effect), were collisionally emplaced to the Trojan cloud, or that they formed in situ, and have simply resided in the Trojan clouds ever since. Such models therefore predict a population of objects strongly concentrated at very low inclinations and eccentricities (e.g. [4,5]), which is quite different to that observed today (e.g. [3,11]). For simplicity, we henceforth refer to these models as 'in situ' formation scenarios.

---

[2] Although the total population of Neptunian Trojans may well exceed that of the asteroid belt by an order of magnitude, or more, it is important to note that the largest members are far smaller than the largest members of the asteroid belt (estimated diameters of less than 200 km versus the ~900 km diameter of the dwarf planet Ceres). This, alone, means that the Neptune Trojan population probably contains less mass than the asteroid belt. The physical nature of the Neptunian Trojans, in addition, is likely very different to the bulk of objects in the asteroid belt, which will exacerbate this difference in total mass. Where the asteroids are typically rocky or metallic, the Neptunian Trojans are most likely primarily icy bodies, similar in constitution to the members of the trans-Neptunian and Centaur populations (and, by extension, to the cometary nuclei sourced from those populations).

More recent studies of the formation of the Solar System (e.g. [6,7]) have invoked the post-formation migration of the giant planets as a process through which the system's small bodies can be dynamically excited to orbits that are significantly "hotter"[3] than those they would originally have occupied. Rather than having formed in situ, upon orbits similar to those they currently occupy, it is now thought most likely that the Neptunian Trojans (like the Jovian Trojans) are a population that was dynamically captured by the planet during its migration outwards through the Solar System [8,9,10,11], and which have since remained trapped in the planet's 1:1 mean-motion resonance over the age of the system [12].

One key prediction that derives from such models is that, if the population truly was captured, rather than having formed in situ, it will consist of objects moving on orbits with a range of dynamical stabilities. Some fraction of the Neptunian Trojan population will be moving on orbits that are less tightly trapped within the 1:1 mean-motion resonance with the planet, and hence will be dynamically unstable on a variety of timescales. Since the Trojans were captured over four billion years ago, those captured to the least stable orbits would be expected to have escaped from the Trojan cloud a long time ago. It seems reasonable to expect, however, that some fraction of the current population should display dynamical instability on timescales of hundreds of millions of years. Indeed, such a mechanism has been invoked by [13,14] to propose that the Neptunian Trojans could well constitute a significant source of fresh material to the dynamically unstable Centaur population[4]. In addition, given that dynamical processes are time-reversible, it is also possible that some fraction (albeit most likely only a small one) of the Trojan population are more recently captured dynamically unstable objects [24]. As such, it is clearly of interest to examine the dynamical behaviour of the individual Neptunian Trojans, to see whether their evolution is in keeping with this prediction.

In this work, we present the results of detailed dynamic simulations of two of the Neptunian Trojans – 2001 $QR_{322}$ and 2008 $LC_{18}$. In the next section, we discuss the case of 2001 $QR_{322}$, the first Neptunian Trojan to be discovered, before moving on to detail the results of a similar, but more detailed, study of the behaviour of 2008 $LC_{18}$.

## 2001 $QR_{322}$

When 2001 $QR_{322}$ was discovered, investigations were made of its dynamical behaviour. Chiang et al. ([2]) considered a small number of test particles based on the observational range of the object's orbital elements were integrated for a period of 1 Gyr. The authors found that the test particles displayed stable behaviour over that period. Marzari, Tricarico & Scholl ([24]) went further, using dynamical simulations to conclude that the object is most likely primordial in nature, with only 10% of their test population of 70 clones escaping from the Trojan cloud in the 4.5 Gyr simulations they carried out. In a more detailed study, Brasser et al. ([25]) found that the majority of clones of the object remained stable in integrations spanning 5 Gyr, although they did note that the nearby $\nu_{18}$ nodal secular resonance (characterised by the libration of $\Omega - \Omega_N$ with time, where $\Omega$ is the longitude of the ascending node and the subscript N refers to Neptune) did result in the object's behaviour within the Trojan cloud being complex in nature.

---

[3] i.e. orbits with significantly higher eccentricities and inclinations than would be expected for objects that formed from a dynamically cold disk of material.

[4] The Centaurs (e.g. [15,16,17]) are a population of dynamically unstable icy bodies whose orbits have perihelia between those of Jupiter and Neptune. The largest Centaurs, (10199) Chariklo and (2060) Chiron, are thought to be between 200 and 250 km in diameter, and it has been estimated (e.g. [16]) that there are approximately 44,000 Centaurs greater than 1 km in diameter currently moving in the outer Solar System. They are widely accepted as the proximate parent population of the short-period comets, which in turn constitute a significant contribution to the impact flux experienced by the Earth [18]. However, the origin of the Centaurs is still the subject of much debate, with suggested parent reservoirs including the Edgeworth-Kuiper belt [19,20], the Scattered Disk [21], the inner Oort cloud [22], and even the Jovian and Neptunian Trojans [13,14,23].

In the time since these studies of the dynamical behaviour of 2001 QR$_{322}$ were carried out, the precision with which the object's orbit is known has increased dramatically, as a result of the arc over which it has been observed growing ever longer. The nominal best-fit orbit for the object has changed somewhat as new observations have been made, but its dynamical behaviour has been neglected, as an ever-increasing number of objects in the outer Solar System have vied for attention. We therefore decided, on the basis of some preliminary trials (detailed in [8]), to revisit the dynamical stability of 2001 QR$_{322}$, taking account of its improved orbital parameters. Thanks to the rapid growth of computing capability over the years since those first studies, it was possible to examine the object's behaviour in far more detail than those previous works.

In order to examine the behaviour of 2001 QR$_{322}$, we used the *Hybrid* integrator within the *n-body* dynamics package *MERCURY* [26] to perform detailed dynamical simulations of almost 20,000 massless test particles, spread across the full 3σ orbital uncertainties in all six of the object's orbital elements, for 1 Gyr. The *Hybrid* integrator within *MERCURY* is a particularly efficient tool for the analysis of orbital evolution, since it offers an excellent compromise between integration speed and the ability to handle close encounters between two bodies. As such, it is both widely used in Solar System astronomy [16,17,23,27,28], astrobiology [29,30,38], and even the study of exoplanetary orbits [31,32,39].

Our integrations of the behaviour of 2001 QR$_{322}$ were based on orbital elements obtained from [33] on 26$^{th}$ January 2009, as shown in Table 2.

| Element | Value | 1σ uncertainty |
|---|---|---|
| *a* (AU) | 30.3023 | 0.008813 |
| *e* | 0.031121 | 0.0003059 |
| *i* (°) | 1.323 | 0.0009417 |
| *Ω* (°) | 151.628 | 0.02328 |
| *ω* (°) | 160.73 | 0.8316 |
| *M* (°) | 57.883 | 0.7818 |
| **Epoch (MJD)** | 54800 | |

*Table 2: The orbital elements, along with their associated 1σ uncertainties, of the nominal best-fit orbit for 2001 QR$_{322}$ as obtained from [33] on 26$^{th}$ January 2009. The values are based on an observational arc of 1450 days, and will doubtless undergo small changes as future observations are made.*

Based on the orbital elements shown in Table 2, and following the procedure established in [16,17], we created a suite of 19,683 test particles, centred on the nominal orbit. These clones were created such that 9 unique values each of *a*, *e*, and *i* were tested, distributed evenly across the ±3σ error ranges for those elements. For each of the 729 (9$^3$) *a-e-i* locations tested, we carried out 27 unique trials, testing three values each for *Ω, ω* and *M*, again spread evenly across the ±3σ error ranges for those elements. The test particles were then followed under the gravitational influence of Jupiter, Saturn, Uranus and Neptune for a period of 1 Gyr, with an integration time-step of 1/3 of a year[5]. As is standard in studies of the dynamics of objects in the outer Solar System, only gravitational perturbations were considered. Non-gravitational forces (such as the Poynting-Robertson and Yarkovsky effects, and potential outgassing from the surfaces of the objects studied) are implicitly assumed to have only a negligible effect on the statistical evolution of the sample, and are therefore not included in our simulations.

Startlingly, given the results of the previous studies into the stability of 2001 QR$_{322}$'s orbit ([2,24,25]), we found that, of an initial swarm of 19,683 test particles, just 7,220 survived for the full 1 Gyr of integration time. Just over 63% of the particles were either ejected to a heliocentric distance of over 1,000 AU or collided with one of the massive bodies (the four giant planets and the Sun). As we discuss in detail in [34], the number of surviving test particles decays in an approximately exponential manner, as a function of time. As such, and following [15,16,17], we can describe the dynamical stability (or lack of it) of 2001 QR$_{322}$ in a statistical sense by means of a "dynamical half-life". Whilst for the truly dynamically unstable members of the Solar System (such as the Centaurs and short-period comets), dynamical half-lives are typically measured in hundreds of thousands, or a few million years, we find that 2001 QR$_{322}$ has a dynamical half-life, based on our results, of 593 million years. The decay of the population of clones of 2001 QR$_{322}$ can be seen in Figure 1.

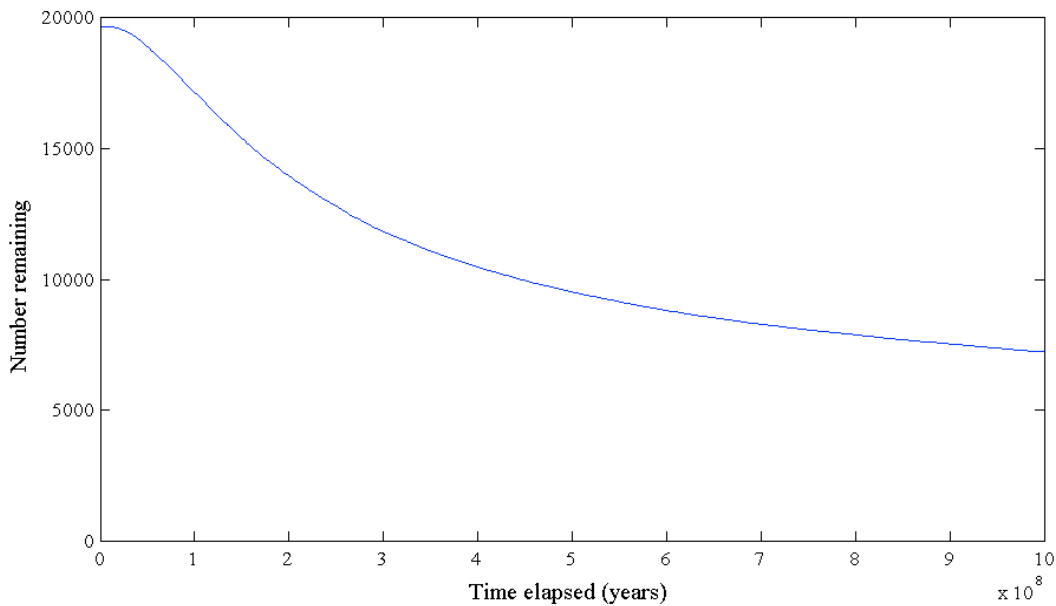

*Figure 1: The decay of the population of clones of 2001 QR$_{322}$ as a function of time.*

---

[5] One of the major innovations in Chambers' *MERCURY* package was the development of his *Hybrid* integrator. That tool has greatly benefitted studies such as this by enabling simulations to be run as quickly as possible whilst maintaining the ability to accurately follow close encounters between objects. When the test particles used in this work are far from any of the massive bodies (Jupiter, Saturn, Uranus and Neptune), the orbital evolution of the particles is calculated using a *symplectic* integrator, with the aforementioned time-step (1/3 of a year). When one of the test particles approaches on of the massive bodies sufficiently closely, however, the evolution is followed instead by a *Bulirsch-Stoer* integrator, which allows the close encounter to be accurately modelled. For further information on the functionality of *MERCURY*, we direct the interested reader to [26].

On the basis of these results it is clear that, although 2001 QR$_{322}$ is a dynamically long-lived object (with a dynamical half-life of around ~2/15$^{th}$ the age of the Solar System), it is not truly dynamically stable on Gyr timescales. However, it is still reasonable to consider that 2001 QR$_{322}$ is an object that formed, or was captured into, the Neptunian Trojan population during the early days of the Solar System, rather than being a more recently captured interloper, given the relatively slow dynamical decay of the population of 2001 QR$_{322}$ clones. Since it seems likely that Neptune would have captured at least ~0.03 Earth masses of material as Trojans like those observed today[6] [8], a sizeable population of Neptune Trojans would exist at the current epoch, even if all those captured objects were trapped on orbits of similar stability to that of 2001 QR$_{322}$.

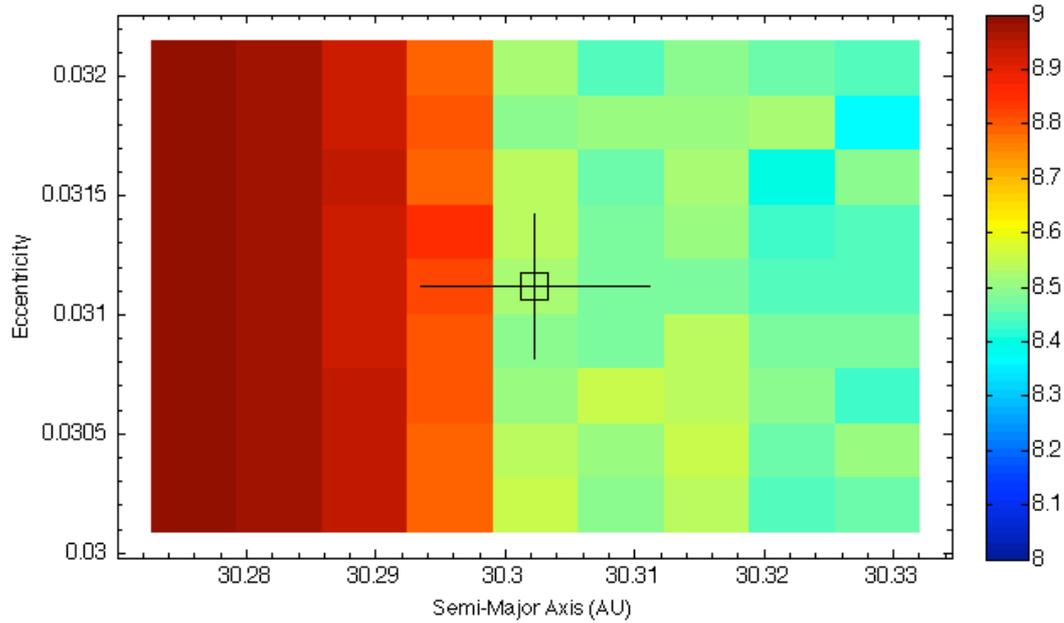

*Figure 2: The dynamical stability of 2001 QR$_{322}$ as a function of its initial semi-major axis and eccentricity. Each square represents the mean lifetime of 243 separate trials, each of which comprised a unique i-Ω-ω solution. It is clear that there is a marked disparity between orbits at semi-major axes greater than 30.29 AU, which tend to be highly dynamically unstable, and those interior to that value, where the orbits are far more stable. The location of the best-fit solution for the orbit of 2001 QR$_{322}$ is shown marked by the square in the centre of the plot, with the associated 1-σ uncertainties in a and e shown by the lines radiating from that point.*

As can be seen in Figure 2, there is a marked difference in the stability of those clones of 2001 QR$_{322}$ that started on orbits with semi-major axis greater than 30.29 AU, and those which started interior to that value. The large area of instability stretching outward from 30.29 AU is the primary reason for the dynamical instability of 2001 QR$_{322}$, and as such merits some further discussion. Upon closer investigation of the behaviour of the clones of 2001 QR$_{322}$ we found that for orbits interior to $a \sim 30.29$ AU the initial libration amplitude of the test particles was typically less than $\sim 60$-$70°$. The majority of such clones survived until the end of the simulations. By contrast, those clones which started at $a > 30.29$ AU had initial libration amplitudes in the range 65-70°, and the great majority of such clones went on to escape from the Trojan region. Although their behaviour superficially appeared similar to that of objects interior to 30.29 AU whilst they were librating as Trojans, it was apparent that their initially slightly greater libration amplitudes allowed them to evolve onto orbits with libration amplitudes of 70-75° on fairly short timescales, at which point their orbits rapidly became unstable. As we describe in more detail in [34], this instability does not appear to be the result of the influence of any secular resonances, nor did the particles experience abrupt changes in orbital or resonant behaviour. It is interesting to note, however, that such large libration amplitudes overlap with the approximate boundary between chaotic and regular motion, according to dynamical diffusion maps for orbits in the vicinity of the Neptunian Lagrange points (e.g. [24,35,36]). It is thought that the broad region of unstable and chaotic resonant motion into which the error-ellipse of 2001 QR$_{322}$'s orbit overlaps might well be the result of a family of complex secondary resonances involving the frequencies of the Trojan librational motion, the near-resonant behaviour of Uranus and Neptune (which lie very close to mutual 2:1 mean-motion resonance), and the apsidal motion of Saturn [36].

---

[6] This estimate is based on the assumption that the disc of material through which Neptune migrated contained of order 30 Earth masses of material, with capture rates as detailed in Table 4 of [8]. We note here that of order 1% of the test particles were captured in the simulations detailed in that work, which suggests that the initial captured population could have contained up to 0.3 Earth masses of material. However, between 80 and 90% of the captured Trojans noted in that work were trapped in horseshoe orbits (librating around both L4 *and* L5 simultaneously), which are inherently less stable than the tadpole-type orbits (in which the Trojan librates around either L4 *or* L5) observed for the eight known Trojans. As such, our cautious estimate assumes just 10% of the captured Trojans were emplaced on tadpole orbits.

## 2008 LC$_{18}$

Minor planet 2008 LC$_{18}$ was the first Trojan to be found librating around Neptune's trailing Lagrange point, L5. The best-fit orbit for 2008 LC$_{18}$ is based on a small number of observations, spread over an arc of just one year[7]. This results in the object having relatively large errors, compared to the best-fit orbits of the other Neptunian Trojans. However, we note that the current scale of the uncertainties on the orbit, as taken from [33] on 23$^{rd}$ August 2011 (as detailed in Table 3), is already comparable to the errors for 2001 QR$_{322}$ at the time of the first dynamical studies of the behaviour of that object [2.24.25]. As such, it seems timely to investigate the behaviour of 2008 LC$_{18}$, the first Neptunian Trojan found librating around the planet's trailing Lagrange point, L5.

| Element | Value | 1σ uncertainty |
|---|---|---|
| $a$ (AU) | 29.9369 | 0.02588 |
| $e$ | 0.083795 | 0.002654 |
| $i$ (°) | 27.569 | 0.003824 |
| $\Omega$ (°) | 88.521 | 0.0007854 |
| $\omega$ (°) | 5.135 | 10.85 |
| $M$ (°) | 173.909 | 12.83 |
| Epoch (MJD) | 55800 | |

*Table 3: The orbital elements of 2008 LC$_{18}$, as obtained from [33] on 23$^{rd}$ August 2011. Note the particularly large errors in the values of ω and M that hold the bulk of the uncertainty in the orbit of this object.*

---

[7] Although 2008 LC$_{18}$ was discovered 3 years ago, it has not been observed since 2009. At the current epoch, it is located within one of the most densely populated star-fields in the night sky, very close to the galactic centre. In [40], we detail unsuccessful attempts to recover 2008 LC$_{18}$ in August 2011 using the 2.3 m telescope at Siding Spring Observatory. The recovery of 2008 LC$_{18}$ will probably require observations using 8m-class telescopes, or a lengthy wait for its orbital motion to carry it away from the direction of the galactic centre.

In order to examine the dynamical behaviour of 2008 LC$_{18}$, we followed the technique described previously, in connection with 2001 QR$_{322}$, and discussed in more detail in [34], to create a suite of clones of the object. As a result of significant improvements in the computational facilities available to us, we were able to study a much larger sample of test particles, and hence obtain better resolution in *a-e-i-Ω* space. Given the size of the uncertainties in the other two orbital elements, we chose to create our suite of test particles purely in *a-e-i-Ω* space. In this case, we created a suite of 25x15x15x11 clones in *a-e-i-Ω*, again spread evenly across the ±3σ error ellipses in each element, for a total of 61875 test particles, which we then followed under the gravitational influence of the Sun, Jupiter, Saturn, Uranus and Neptune for a period of 1 Gyr, again using the *Hybrid* integrator within *MERCURY* [26].

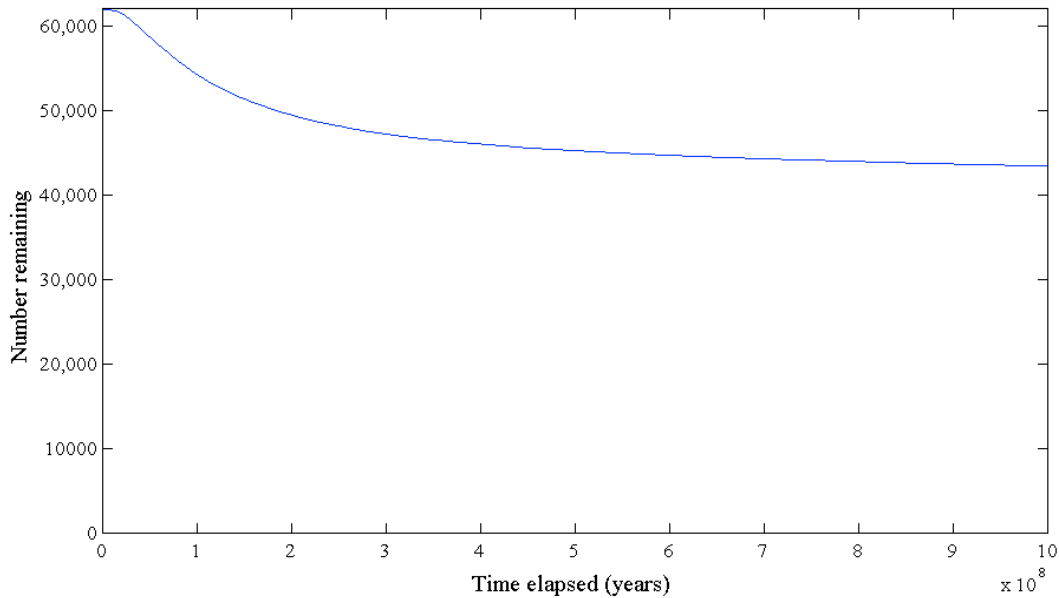

*Figure 3 – The decay in the population of clones of 2008 LC$_{18}$ as a function of time.*

It is immediately apparent from inspection of Figure 3 that a sub-population of order 25% of the total number of clones of 2008 LC$_{18}$ are dynamically unstable, whilst the bulk of the remainder are moving on far more stable orbits. As such, the resulting curve is best fit by the combination of the rapid exponential decay of the unstable component combined with the very slow decay of the stable component. To see whether this behaviour is linked to the error-ellipse for 2008 LC$_{18}$ lying in the vicinity of a boundary between stable and unstable orbital element phase space, we again examined the variation of the object's lifetime as a function of semi-major axis and eccentricity. As can be seen from Figure 4, the orbit of 2008 LC$_{18}$ lies close to the boundary between orbits with high dynamical stability (outwards of ~29.1 AU) and those that are extremely dynamically unstable (inwards of ~29.0 AU), with a narrow strip of moderate stability separating the two. Whilst the boundary between stable and unstable regions lay within just 1σ of the nominal best fit orbit for 2001 QR$_{322}$, it instead lies just beyond that range for 2008 LC$_{18}$, which explains why the unstable component of the overall population is that much smaller (as seen in Figure 3). Nevertheless, it is interesting that this object, too, has the potential to turn out to be a dynamically unstable Trojan.

Unlike 2001 QR$_{322}$, where the overall stability was such that it seems reasonable to assume a primordial origin for the object, the situation is less clear-cut for 2008 LC$_{18}$. Given the vast disparity in stability between the stable and unstable orbits, two scenarios seem reasonable for the object. If future study refines the orbit such that the full 3σ error range lies within the stable regime, beyond 29.91 AU, then it seems likely that 2008 LC$_{18}$ is a dynamically stable, and hence primordial, Neptune Trojan. On the other hand, if such study reveals that the object lies sunward of 29.90 AU, then the dynamical lifetimes observed in that region are sufficiently short that it is challenging to believe the object is truly primordial (with typical mean lifetimes of order 100 or 200 million years). As such, it must be assumed that the object only relatively recently moved onto its current orbit. Either it is a recent temporary capture (such as those described by [23]), or something has happened to shift it from a more tightly bound Trojan orbit (such as a recent collision with another trans-Neptunian object). In either case, the object would clearly be an interesting target for further study, both observational and theoretical.

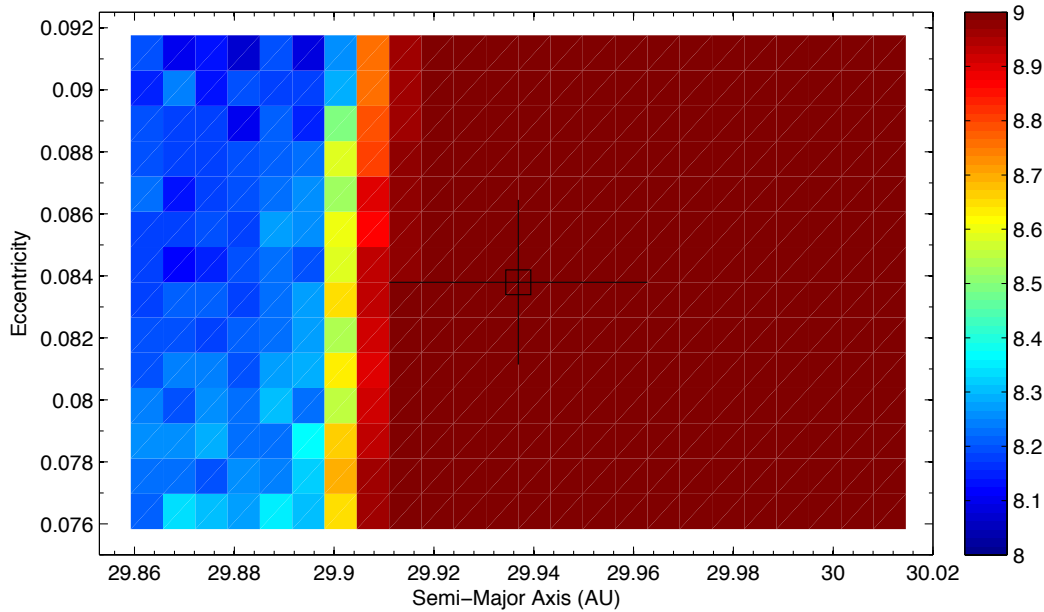

Figure 4: The dynamical stability of 2008 $LC_{18}$, as a function of semi-major axis and eccentricity. As was the case for 2001 $QR_{322}$, discussed above, the orbit of 2008 $LC_{18}$ appears to lie close to a sharp boundary between dynamically stable and dynamically unstable orbits. Each box in the figure shows the mean lifetime of 165 separate trials, each with a unique combination of i and Ω.

## Conclusions

As the most recently discovered family of objects in our Solar System, and the least well studied, the Neptune Trojans are a fascinating test-bed for models of planetary formation and evolution. The current favoured explanation for the origin of Neptune's Trojans is that they were captured during the migration of the planet. As such, it would be expected that Trojans were captured on a wide variety of orbits, with a wide variety of dynamical stabilities, and that therefore at least some of the Neptunian Trojans should display dynamical instability on timescales of hundreds of millions or even a few billion years.

In this work, we present the results of detailed dynamical simulations of the first Neptunian Trojan discovered, 2001 $QR_{322}$, and the first found librating around the planet's trailing Lagrange point, 2008 $LC_{18}$. In turns out that, for both objects, the error ellipse around their best-fit orbits spans both dynamically stable and dynamically unstable regions. In the case of 2001 $QR_{322}$, whose orbit is now relatively well-constrained, the unstable-stable boundary lies within 1σ of the best fit orbit, whilst for the less well-constrained 2008 $LC_{18}$, the boundary lies just beyond 1σ from the best fit orbit.

In coming years, it is imperative that further observations of these objects are carried out, in order to determine their true dynamical nature. It is possible that one, or even both, objects are dynamically stable – either being the unstable left-overs of a once larger captured population, or objects that were captured from the Centaur population in the relatively recent past.

Our results for 2001 QR$_{322}$ also highlight the need to revisit the orbital behaviour of previously studied objects once their orbits have become significantly better constrained. The dynamical behaviour we detail for that object is in stark contrast to that noted in the first studies carried out for the object, almost a decade ago. That disparity is a direct result of the fact that, as the observational arc for the object has grown, the best-fit elements for its orbit have shifted significantly, such that it now samples a noticeably different regime within Neptune's leading Trojan cloud.

## Acknowledgements

JH acknowledges the financial support of the Australian Research Council, through the ARC Discovery Grant DP774000. The authors wish to thank the two referees of this work, whose suggestions helped to improve the flow and clarity of the article. They simulations carried out in this work were performed using the *n*-body dynamics package *MERCURY*, written by Dr. John Chambers [26], and were based on orbits obtained from the *AstDys* website (http://hamilton.dm.unipi.it/astdys/), which is maintained by funding from the University of Pisa, the Agenzia Speziale Italiana, the Astronomical Observatory of Belgrade and the University of Valladolid, and uses observations published by the IAU Minor Planet Center (http://www.minorplanetcenter.net/iau/mpc.html).